\documentclass[runningheads, dvipsnames]{llncs}

\usepackage{amssymb}
\usepackage{booktabs}
\usepackage{multirow}
\usepackage[inline]{enumitem}
\usepackage{subcaption}

\usepackage{wrapfig}

\usepackage[numbers]{natbib}

\usepackage{xcolor}
\usepackage{graphicx}
\let\llncssubparagraph\subparagraph
\let\subparagraph\paragraph
\usepackage[compact]{titlesec}
\let\subparagraph\llncssubparagraph

\begin{document}
\title{Behavioural Effects of Agentic Messaging}
\subtitle{A Case Study on a Financial Service Application }

\author{Olivier Jeunen \and Schaun Wheeler}
%\author{}\institute{}
\institute{aampe\\
\email{\{firstname\}@aampe.com}}
%\email{firstname@aampe.com}

%
\maketitle              % typeset the header of the contribution
\begin{abstract}
Marketing and product personalisation provide a prominent and visible use-case for the application of Information Retrieval methods across several business domains.
Recently, agentic approaches to these problems have been gaining traction.
This work evaluates the behavioural and retention effects of agentic personalisation on a financial service application's customer communication system during a 2025 national tax filing period.
Through a two month-long randomised controlled trial, we compare an agentic messaging approach against a business-as-usual (BAU) rule-based campaign system, focusing on two primary outcomes: unsubscribe behaviour and conversion timing.
Empirical results show that agent-led messaging reduced unsubscribe events by 21\% ($\pm 0.01$) relative to BAU and increased early filing behaviour in the weeks preceding the national deadline.
These findings demonstrate how adaptive, user-level decision-making systems can modulate engagement intensity whilst improving long-term retention indicators.
\end{abstract}

% \maketitle

\section{Introduction \& Motivation}
Consumer businesses seek to keep their customers engaged, optimising communications for their incremental effect on key user behaviours.
An effective use of contextual information as well as every user's personal interaction history are imperative to success~\cite{Kumar2018}. 
Recent work has shown that \emph{agentic} messaging approaches---built on foundational ideas from econometrics, causal inference and contextual bandits~\cite{joachims2021recommendations,CONSEQUENCES2022}---provide a promising avenue to tackle these problems in practice~\cite{Abboud2025}.

Our work summarises a case study of such an approach on a financial service application, designed to simplify income tax filing through guided mobile and web interfaces.
Its core mission---to make tax filing accessible to individuals who would otherwise skip or overpay---relies heavily on sustained engagement and trust.
The conventional approach to campaign orchestration in Customer Relationship Management (CRM) is entirely based on iteratively optimised messaging rules on pre-defined customer segments~\cite{Tynan1987}.
The key analytic question for the application is whether an agentic approach is successful in driving more consistent user behaviour and reducing message fatigue: leading users to act earlier, with fewer opt-outs, within the real-world constraints of a seasonal filing cycle.
This question sits at the intersection of applied causal inference and product design.
Across the marketing technology ecosystem, automated personalisation is often conflated with A/B-testing at scale to iterate on rule improvement---even though it is known to be a provably suboptimal decision-making methodology~\cite{Garivier2016}.

The agentic approach reframes this as a sequential decision-making problem---each message an experiment, each user an evolving context. The goal is to learn adaptive policies that generalise, not just to optimise one marketing campaign.
\subsection{An Agentic Approach}
The system evaluated here extends the agentic infrastructure described by Abboud et al. \cite{Abboud2025}.
In this setup, autonomous agents act as decision-makers over multiple dimensions of communication:
\begin{enumerate*}[label=(\roman*)]
    \item When to contact a user,
    \item through which channel (e-mail, push, in-app),
    \item with what message variant (tone, incentive, or call-to-action).
\end{enumerate*}
Each agent continuously updates its beliefs about the effect of a potential message on downstream user actions using Bayesian Thompson sampling)~\cite{Chapelle2011,Jin2023,Russo2018}, under a Difference-in-Differences (DiD) framework for incremental impact estimation~\cite{Athey2006}.
The control group continued to receive manually orchestrated messages following fixed campaign schedules and segment rules, whilst the agentic group received dynamically timed and personalised interventions learnt over time.
Both systems operated on the same event data infrastructure, with consistent eligibility and suppression rules.
Note that the rule-based BAU system reflects years of accumulated human judgment—dozens of segment definitions, scheduling choices, and message variations shaped through trial, habit, and intuition. Each adjustment represents real expertise, yet the process leaves no formal record of how or why those parameters evolved. The result is a sophisticated but opaque construct: a system that works largely because people have made it work, but one whose internal logic can't easily be traced, tested, or learned from. It stands as both a testament to coordinated human effort and a reminder of how much learning disappears when optimisation lives only in people's heads.
The agentic treatment, however, learns from scratch and evolves continually throughout the experiment period.
\section{Experimental Setup and Design}
The goal of this analysis is to isolate the causal impact of agentic personalisation.
We describe the experiment setup, data sources, and inference methods used to distinguish genuine behavioural effects from coincidental trends.
The evaluation followed a randomised controlled trial (i.e. an A/B-test~\cite{kohavi2020trustworthy,Jeunen2025}) during a 2025 national tax filing season.
Users were randomly assigned to either:
\begin{description}
\item[\textbf{BAU} (Control):] Campaigns following traditional segmentation and cadence.
\item[\textbf{Agentic} (Treatment):] Adaptive, agentically managed messsaging strategies.
\end{description}
Groups were split equally among $6.4$ million users.
Event stream data was collected for messaging, unsubscribes, intent and conversion events.

Two outcomes were analysed:
\textbf{Unsubscribe Rate}: frequency of e-mail unsubscription events, disaggregated by sender (BAU vs Agentic).
This message source distinction avoids confounding between mixed event streams.
\textbf{Filing Timing}: the daily cumulative proportion of users who submitted their tax return.

For unsubscribe comparisons, we computed the relative difference between BAU-triggered and Agent-triggered unsubscribe rates within the active campaign window.
Confidence intervals were derived from standard errors of binomial proportions.
For behavioural timing, we calculated week-level submission rates and the differential trajectory between treatment and control groups over the filing period.
The difference-in-differences estimator used by the agentic treatment, captured and incentivised incremental acceleration toward submission relative to the common deadline.

All results were validated for start-date sensitivity and group assignment stability, following data alignment and sample ratio mismatch checks.

\subsection{Agentic Messaging Leads to Fewer Unsubscription Events}
\begin{figure}[h]
    \centering
    \begin{subfigure}[h]{.48\textwidth}
    \includegraphics[width=\linewidth, trim={0cm 0.5cm 0cm 1.3cm}]{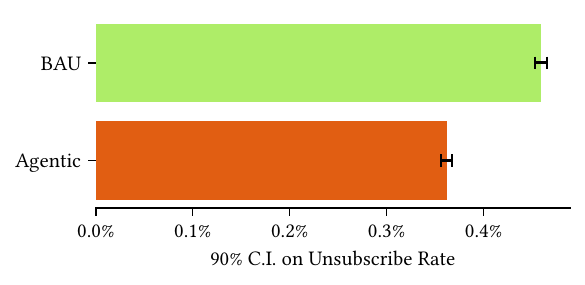}
    \caption{Unsubscribe rate per treatment.}\label{fig:1a}
    \end{subfigure}~
    \begin{subfigure}[h]{.5\textwidth}
    \includegraphics[width=\linewidth, trim={0cm 0.5cm 0cm 1.3cm}]{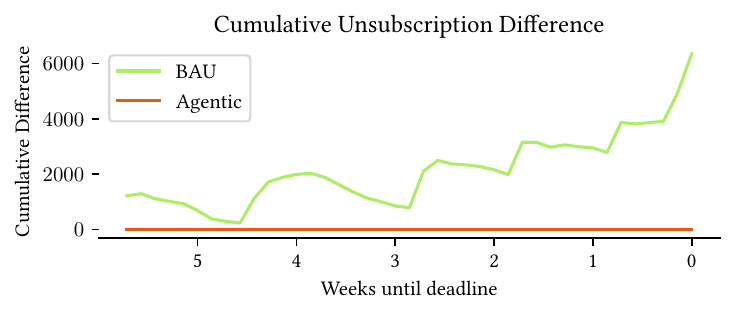} 
        \caption{Cumulative unsubscriptions per group.}\label{fig:1b}
    \end{subfigure}
    \caption{Empirical results for unsubscription rates in the A/B-test.}
    \vspace{-5ex}
\end{figure}

We report quantitative differences between the agentic and control groups for unsubscribe behaviour, evaluating the statistical strength and practical relevance of those effects.
Across the campaign period, Agent-initiated messages yielded 21\% fewer unsubscribe events than rule-based messages, controlling for exposure. The 90\% confidence interval for the relative lift was ($ –0.212 \pm 0.0098$), yielding an extremely low $p$-value to reject a no-difference null hypothesis (Fig.~\ref{fig:1a}).

The reduction was not due to reduced send volume: we validated that messaging frequencies were stable and consistent in the Agentic treatment.
Instead, the pattern reflects behavioural differences in user response. Cumulative analyses show that unsubscribe rates from rule-based messages spiked periodically with campaign bursts, with the Agentic unsubscribe curve rising more gradually and stabilising earlier. The cumulative differential widened steadily across the eight-week horizon, emphasising compounding effects over time (Fig.~\ref{fig:1b}).

This indicates that agentic scheduling---through selective targeting and adaptive pacing---mitigated message fatigue while maintaining communication reach.

\subsection{Agentic Messaging Nudges Customers to Convert Earlier}
\begin{figure}[h]
    \centering
    \includegraphics[width=0.9\linewidth, trim={0cm 0.7cm 0cm 1.5cm}]{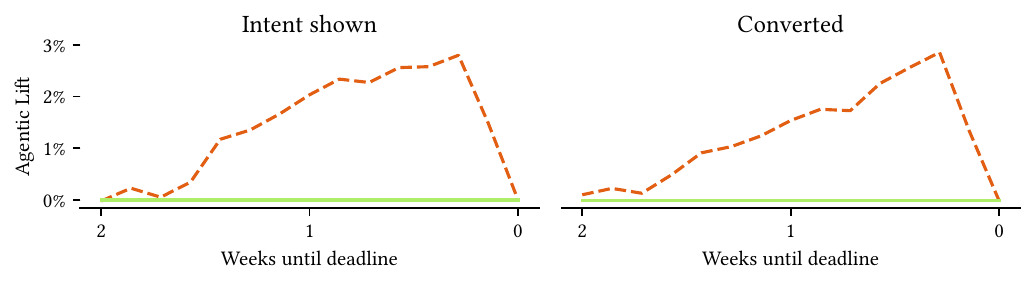}
    \caption{Relative increase in event frequencies in the A/B-test.}
    \vspace{-5ex}
\end{figure}
In the weeks preceding the national tax filing deadline, users who were exposed to agentic messaging exhibited higher submission rates than those in the BAU group.
The effect was most pronounced between three and one weeks before deadline, consistent with the hypothesis that more tailored message timing enhances action readiness.
This effect holds across the funnel---for users who convert, but also those who perform upper-funnel events showing intent.

Although both groups ultimately converged by the deadline, the temporal distribution of activity differed: the agentic cohort filed earlier on average, flattening the peak load seen in control.
This behavioural shift holds operational and psychological significance---suggesting that personalised pacing helps users complete complex administrative tasks sooner, not just more often.

For a use-case like tax filing where every conversion entails some operational load to the business, the observed temporal redistribution of conversion events brings desirable practical benefits that enhance scalability and cost-efficiency.
\section{Discussion, Conclusions \& Outlook}
The unsubscribe and timing effects together illustrate the central design principle of agentic personalisation: learning when \textit{not} to send.
Traditional marketing optimisation tends to reward high message throughput; agentic systems instead optimise expected incremental outcomes under uncertainty.
The observed 21\% reduction in unsubscribes corresponds to a statistically robust decline in negative engagement events---a critical metric in retention-heavy application domains. Because unsubscribes are irreversible, even small percentage differences represent substantial long-term audience preservation, and effects compound over time.
The early-filing uplift demonstrates how adaptive sequencing can align behavioural nudges with individual readiness, effectively reshaping aggregate funnel dynamics without increasing message volume.

From a product perspective, these results show measurable efficiency gains: more engagement from fewer total interventions, with reduced attrition risk. For a seasonal business such as tax filing, spreading submissions earlier in the cycle also reduces support bottlenecks and infrastructure strain.
Naturally, the data carry caveats. Unsubscribe and submission events are only partially observable; a subset of users triggered multiple unsubscribes, possibly due to client-side retriggers or category-level preferences. These artefacts do not invalidate the observed directionality but limit precision at the individual level.
The agentic advantage should therefore be interpreted as a robust population-level effect rather than an exact per-user estimate.

We highlight three lessons for deploying agentic systems at scale:
\begin{enumerate}
    \item Event integrity is non-trivial. Even mature data stacks exhibit ambiguity around what constitutes a user action versus a logging artifact. Reducing this ambiguity is critical for reproducibility.
    \item Adaptive restraint is as valuable as adaptive persuasion. The most effective agentic policies often emerged from not sending---recognising when further nudges would add noise rather than value.
    \item Temporal redistribution is a valid success metric. Shifting engagement earlier in time can yield organisational benefits independent of aggregate conversion rates. Agentic systems allow such temporal learning to occur autonomously.
\end{enumerate}
Viewed collectively, these findings support a broader thesis: agentic personalisation enables marketing to move from rule execution to behavioural learning---an architecture that not only reacts to user preferences but evolves with them~\cite{Wheeler2025}.

This work represents an in-depth case study for one specific application domain.
Nevertheless, the methodology and framework we evaluate is agnostic to the use-case.
This emboldens our belief that these results are not limited to the financial services domain, but hold a promise for generalisability across various industries and application areas.

\section*{About the presenter}
Olivier Jeunen is a Principal Research Scientist at Aampe, after holding positions at Amazon, Spotify, and Meta, among others. His research focuses on machine learning for decision-making, marrying probability-theoretic ideas with applications in recommendation and personalisation. This has led to 50\textsuperscript{+} publications, two Best (Student) Paper Awards, and multiple recognitions as an Outstanding (Senior) PC Member. He co-organises the CONSEQUENCES workshop series, and served as an Industry Chair for ECIR '24 and RecSys '25.

\section*{About the company}
Aampe is a science-driven start-up building agentic infrastructure to reconfigure and reorient how business communicate with their users.
Its agents model individual user behaviour and preferences to optimise message content, sequencing, and timing in real-world applications. Founded by researchers and engineers from leading academic and industry institutions, Aampe's work bridges applied machine learning and behavioural science, advancing data-efficient, adaptive methods for personalisation beyond traditional experimentation and A/B-testing frameworks.

\bibliographystyle{splncs04}
\bibliography{references}

\end{document}